 \documentclass[final,3p,times,twocolumn]{elsarticle}

\usepackage{graphicx}
\usepackage{amssymb}

\usepackage{lineno}

\usepackage{rotating} 

\biboptions{square,sort&compress}

\journal{Scripta Materialia}

\begin{document}

\begin{frontmatter}


\title{The lure of ice-templating: recent trends and opportunities for porous materials}




\author{Sylvain Deville}
\ead{sylvain.deville@saint-gobain.com}

\address{Ceramics Synthesis and Functionalization Lab, UMR 3080 CNRS/Saint-Gobain, Cavaillon, France.}

\begin{abstract}
Ice-templating is a simple materials processing and shaping route where the growth of crystals from a solvent template the porosity. Although the principles of ice-templating were established a long time ago, it has lured considerable attention over the past 5 years to become a well-established processing route for porous materials of all kinds. In this viewpoint, I summarize the current status and recent trends of ice-templating studies. I then review the unique assets of ice-templating and propose a roadmap of the most promising or pressing questions and opportunities in this domain.
\end{abstract}


\end{frontmatter}


\section{Introduction} 
\label{sec:introduction}

When a suspension of particles is frozen, the growing crystals can repel and concentrate the particles in the space between them, until the ice eventually invades the inter-particle space~\cite{Deville2017}. After elimination of the crystals, the arrangement of concentrated particles in the frozen structure is maintained and the pores are a replica of the crystals. The process is simple and versatile. It has thus been tested with all kinds of materials, from ceramics to metals, polymers, carbon materials such as graphene or nanotubes, and composites thereof. During the past 10 years, the status of ice-templating has evolved from a new processing route to an established one. With hundreds of ice-templating papers published, we have now a good overview of what can be achieved and what makes ice-templating unique, in terms of both process and materials. This viewpoint thus aims to discuss the future of ice-templating and highlight areas that probably deserve a particular attention in the future.


\section{Where do we stand today ?} 
\label{sec:where_do_we_stand_today_}

The interest for ice-templating was mostly driven by the anisotropy of the structure and sometimes of the properties. The templating mechanisms by the ice crystals are versatile and have been demonstrated for all classes of materials. Matter, whereas found as particles, monomers, or more generally objects, is segregated by the growing crystals. All ice-templated materials have thus a similar morphology, independently of their nature.

The objective of ice-templating has been, for a long time, to develop macroporosity. The templated pores are usually macropores, with a variety of morphologies. With the usual cooling rates reported (1 to 10K/min), macropores in the 5--50~$\mu m$ range are obtained~\cite{Deville2015a}. However, many studies also demonstrated how to obtain micro- and mesoporosity~\cite{Tamon2015}. Such techniques can be combined with the usual ice-templating routes to obtain hierarchical porosity.

\begin{table*}
\resizebox{\textwidth}{!}{%
  \begin{tabular}{lp{8cm}p{8cm}}
  Properties & Materials considered & Applications \\
  \hline
  Porosity & Ceramics, polymers & Tissue engineering, filtration, membranes, thermal insulation, acoustic adsorption \\
  Mechanical strength & Ceramics, metals & Filtration, tissue engineering  \\
  Toughness & Ceramics, metal/ceramic composites & Armour, impact resistance, high performance structural materials \\
  Specific surface area & Carbon, silica, alumina, titania, graphene, chitosan, nanocellulose & Catalysis, adsorption  \\
  Adsorption & Carbon, cellulose, chitosan, graphene, graphene oxide, copper, clay, silica, collagen & Adsorption of gas and heavy ions \\
  Permeability & Alumina, silicon carbide, silicon nitride, collagen & Filtration, catalysis  \\
  Conductivity & Graphene, carbon nanotubes & Batteries, supercapacitors, sensors  \\
  Dielectric properties & Silicon nitride, SiAlON, lead zirconate, rare-earth silicate  & Microelectronics  \\
  Piezoelectric properties & Barium titanate, PZT, PZN &   \\
  Capacitance & Graphene, graphene oxide, carbon nanotubes, barium titanate & Batteries, supercapacitors \\
  Permittivity & PZT, barium titanate & Dielectrics, microelectronics  \\
  Thermal conductivity & Silica, rare-earth silicate, alumina, zirconia, silicon nitride, mullite, clay, carbon nanotubes & Thermal insulation, flame-retardancy, thermal energy storage, thermoresponsive materials  \\
  \end{tabular}}
  \caption{Main properties and applications targeted for ice-templated materials.}
  \label{tbl:properties}
\end{table*}

The porosity content can be anywhere in the 0--100\% range. On the extreme porosity content side, the main limitation is the strength of the material that we can handle. Such extreme porosity can be a limitation for ceramics or metals, but has been an advantage for polymeric and carbon-based materials, with the possibility discussed below of ice-templating aerogels with extreme porosity content. In the low porosity content range, the development of macroporosity is limited by the ability of ice crystals to repel particles in a highly-concentrated suspension. Above a solid loading threshold around 55--60~vol.\%, the crystals grow directly in the interparticle space, no macroporosity is thus obtained.

Many solvents have been successfully tested, including camphene, tertbutyl-alcohol (TBA), dioxane, naphtalene, cyclohexane, DMSO, and a few others~\cite{Deville2017}. Many of these solvents were used for polymers which are not water-soluble. However, the pore morphology that is sought can also dictate the choice of solvents. 


We also have now a good idea of the typical processing time and size limitations. Because of practical and physical constraints, we do not have a total freedom in terms of sample dimensions/pore size combinations. With freeze-front velocity typically in the 5--50~$\mu m/s$ range, the time required to freeze samples can vary from a few seconds (for thin films or microparticles) to several hours (for thick samples). Besides patience, the control of the freeze front velocity and morphology over long periods (hours) can be problematic. In most situations, one can freeze 1~cm samples within 10~min or so.

Ice-templating is mostly material-agnostic, all kinds of materials have thus been ice-templated. If the structural properties were the main interest for a long time, the last few years have seen many studies focused on other physical and functional properties (Tab.~\ref{tbl:properties}). The range of properties and applications that can be targeted is extremely broad.

Finally, we have a fair understanding of the fundamentals of the process, albeit mostly experimental. For the most part, we know which levers of the process should be pulled to control the microstructure, architecture, and properties of ice-templated materials. However, it is still difficult to predict the precise behavior of the system. A first screening of parameters (solid loading, cooling rate, etc\ldots) must first be done on any new system (material, powder).


\section{Recent trends} 
\label{sec:recent_trends}

Three important trends have been observed in the last 2-3 years: (1) ice-templated metals (2) ice-templated aerogels, and (3) the use of ice-templating to trigger self-organization or self-assembly.

\subsection{Ice-templated metals} 
\label{sub:ice_templated_metals}

Metals are the latest class of ice-templated materials. This delay is related to a few specificities of metals that make ice-templating more difficult:
\begin{itemize}
  \item the density of metals is usually higher than that of polymers and ceramics. Settling issues must be fixed~\cite{Driscoll2011a,Chino2008}.
  \item the availability of metal powders of suitable particle size (submicronic), which at the same time exhibit a limited reactivity for the solvent. The almost instantaneous oxidation of metals particles in water makes their use difficile. Metal oxide powders, reduced before or during sintering, can thus be used~\cite{Plunk2017}. Alternatively, metal precursors (such as $TiH_2$) have also been ice-templated~\cite{Yook2008}, followed by a reducing treatment.
\end{itemize}

\begin{figure*}[htbp]
\centering
\includegraphics[width=16cm]{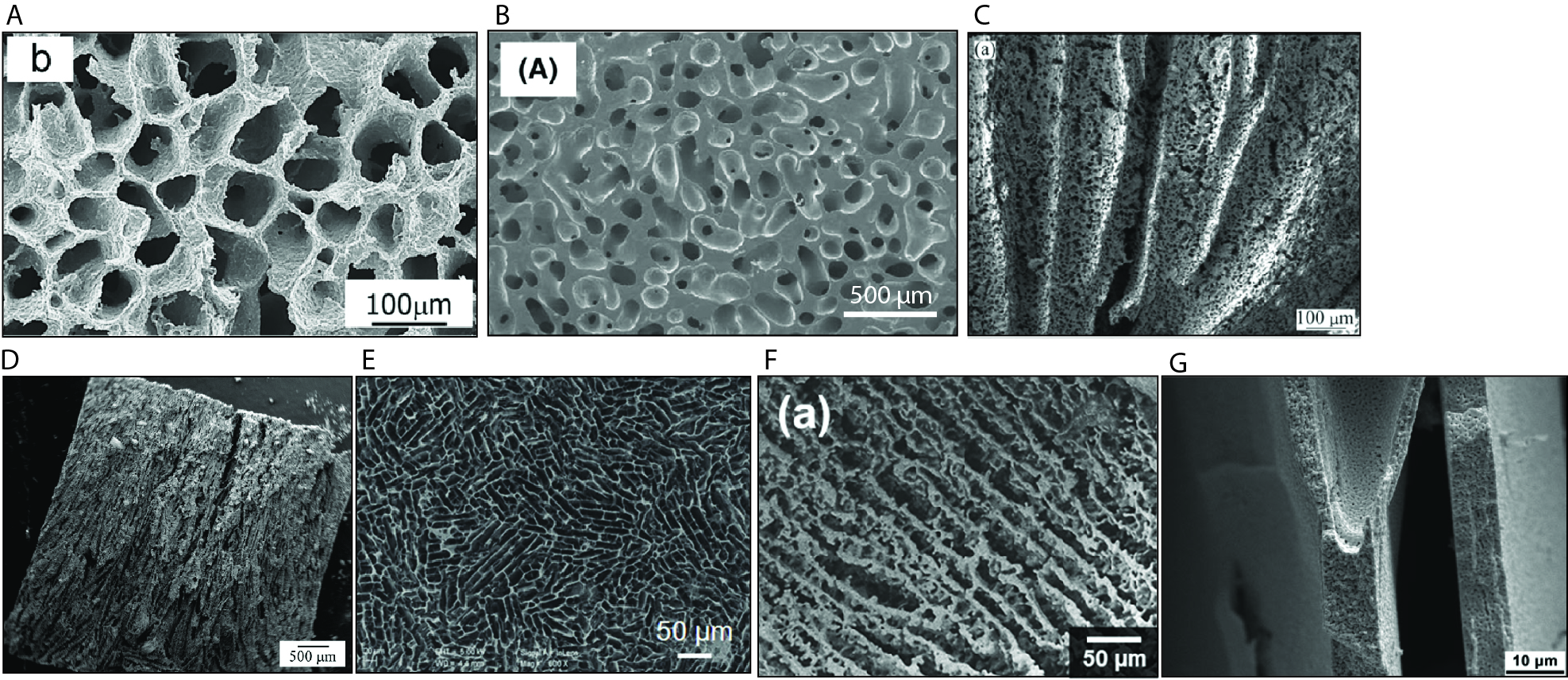}
\caption{Ice-templated metals. (A) to (G): Ni, Ti, Cu, Mo, Ag, Fe, W. Credits: (A) Reprinted from Materials Letters, 153, M. Fukushima, Fabrication of highly porous nickel with oriented micrometer-sized cylindrical pores by gelation freezing method, p.99--101, \copyright~Copyright (2015), with permission from Elsevier (B) Reprinted from Materials Letters, 62(30), S. Yook, Porous titanium (Ti) scaffolds by freezing TiH$_2$/camphene slurries, p.4506--4508, \copyright~Copyright (2008), with permission from Elsevier (C) Reprinted from Transactions of Nonferrous Metals Society of China, 22, S. Oh, Microstructure of porous Cu fabricated by freeze-drying process of CuO/camphene slurry, p.688--691, \copyright~Copyright (2012), The Nonferrous Metals Society of China. Published by Elsevier Ltd. All rights reserved. (D) Reprinted from Materials Letters, 139, S. Oh, Freeze drying for porous Mo with different sublimable vehicle compositions in the camphor-naphthalene system, p.268--270, \copyright~Copyright (2015), with permission from Elsevier (E) Reprinted from H. Gao \emph{et al.}, Macroscopic free-standing hierarchical 3D architectures assembled from silver nanowires by ice templating, Angewandte Chemie \copyright~Copyright 2014 WILEY-VCH Verlag GmbH \& Co. KGaA, Weinheim (F) Metallurgical and Materials Transactions A, Processing, Microstructure, and Oxidation Behavior of Iron Foams, 47(9), 2016, p.4760--4766, H. Park \emph{et al.}, \copyright~Copyright 2016, The Minerals, Metals \& Materials Society and ASM International,  With permission of Springer (G) Reprinted from A. R\"othlisberger \emph{et al.}, Synthesis, structure and mechanical properties of ice-templated tungsten foams, Journal of Materials Research, 31, 6, p.753--764, reproduced with permission.}
\label{fig:metals}       
\end{figure*}

Many metals have now been ice-templated (Fig.~\ref{fig:metals}): Cu~\cite{Tang2013a}, Au~\cite{Freytag2015}, Fe~\cite{Park2016b}, Mo~\cite{Oh2014a}, Ni~\cite{Jo2016}, Sn~\cite{Bang2015}, Ag~\cite{Gouws2015}, Ti~\cite{Jenei2016}, W~\cite{Rothlisberger2016}, or stainless steel~\cite{Driscoll2011a}. Ice-templating is a great alternative to the current routes available to obtain porous metals with unidirectional porosity. Now that the initial issues associated to metals have been fixed, we can expect more systematic assessments of their properties and potential.

\subsection{Ice-templated aerogels} 
\label{sub:ice_templated_aerogels}

\begin{figure*}[htbp]
\centering
\includegraphics[width=16cm]{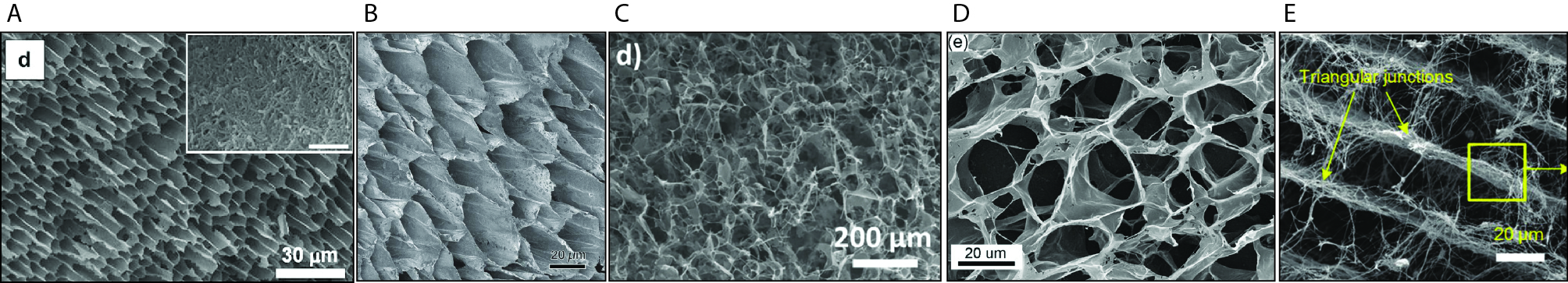}
\caption{Ice-templated aerogels: (A) Carbon nanotubes (B) Nanocellulose/carbon nanotubes (C) Manganese dioxide (D) Graphene (E) Biomass-derived carbonaceous nanofibres. Credits: (A) Reprinted from Polymer, 50(13), S. Kwon, Multiwalled carbon nanotube cryogels with aligned and non-aligned porous structures, p.2786--2792, \copyright~Copyright (2009), with permission from Elsevier. (B) Reprinted with permission from M. M. Hamedi, \emph{et al.}, Highly Conducting, Strong Nanocomposites Based on Nanocellulose-Assisted Aqueous Dispersions of Single-Wall Carbon Nanotubes, ACS Nano, 8, 2467--2476 (2014). \copyright~Copyright (2014) American Chemical Society. (C) Reproduced (in part) from~\cite{Xu2016b} with permission of The Royal Society of Chemistry (D) Reprinted from J. Sun \emph{et al.}, Controllable Fabrication of Transparent Macroporous Graphene Thin Films and Versatile Applications as a Conducting Platform. Adv. Funct. Mater. 25, 4334--4343 (2015). \copyright~Copyright (2015) WILEY-VCH Verlag GmbH \& Co. KGaA, Weinheim (E) Reprinted from Y. Si \emph{et al.}, Ultralight Biomass-Derived Carbonaceous Nanofibrous Aerogels with Superelasticity and High Pressure-Sensitivity. Adv. Mater., 28(43), p.9512--9518 (2016). \copyright~Copyright (2016) WILEY-VCH Verlag GmbH \& Co. KGaA, Weinheim.}
\label{fig:aerogels}       
\end{figure*}

Because the solubility limit of pretty much anything in ice is extremely low, it is possible to start with an extremely dilute suspension and yet concentrate matter between the growing crystals. This unique feature has been used to ice-template a variety of aerogels (Fig.~\ref{fig:aerogels}): carbon nanotubes~\cite{Dong2015,Zhao2015a,Li2016b}, nanocellulose~\cite{Hamedi2014}, carbon nanotubes/chitosan composites~\cite{Yan2014a}, or manganese dioxide~\cite{Xu2016b}. Most of the work was nevertheless concentrated on graphene~\cite{Lin2016} and graphene oxide~\cite{Liu2016a,Shao2016}, when it was realized that ice-templating could be used to obtain bulk macroporous graphene with dimensions (centimeter) of practical sizes. Aerogels with a density as low as 0.14$mg/cm^3$~\cite{Si2016} and 0.16~$mg/cm^3$~\cite{Sun2013} have been reported, and there is no reason why we could not obtain even lower densities, starting with even more dilute suspensions. These aerogels have been considered for a variety of applications such as sensors, catalysis, anodes for fuel cells, battery, or super capacitors, adsorption, or even tissue engineering. Centimeter-size, macroporous graphene with a wall thickness as small as 1.2~nm can be obtained~\cite{Shao2016}.

\subsection{Self-organisation, self-assembly} 
\label{sub:beyond_simple_templating_self_organisation_self_assembly}

The systems investigated for a long time were powders with a broad particle size distribution and no specific morphology. It was recently realized that the concentration increase induced by the freeze-front could be used to trigger self-organization or self-assembly. The objective is thus not to just template the porosity, but also to create particular organizations of objects (particles, micelles) in the frozen structure. Two main ideas have been developed. The first idea is to use anisotropic objects such as platelets or fibers. If properly controlled, the freeze front can align and orient the anisotropic objects, and thus induce preferential orientations in the structure~\cite{Ghosh2016,Martoia2016}, but also create lightweight, percolating fiber scaffolds (SiC, cellulose)~\cite{Ferraro2016,Dash2012}. The alignment of objects can be used later to obtain specific morphologies and thus microstructures~\cite{Bouville2014a} and improves structural or functional properties.
The increase of concentration between the ice crystals can also trigger self-assembly. This was reported with block copolymers, which can self-assemble into a variety of organized structures such as micelles when their concentration increases~\cite{Dhainaut2014,Albouy2016}. These micelles can then self-assemble into more complex configurations. Such self-assembly has been heavily investigated with evaporation, and its application to freezing is recent. This self-assembly has been used to elaborate porous materials with a hierarchical porosity: an organized mesoporosity templated by the micelles, combined with the usual ice-templated macroporosity. 



\section{The lure of ice-templating: what makes it unique ?} 
\label{sec:what_makes_ice_templating_unique_}

The first and maybe most unique feature of ice-templating is the extreme increase of concentration induced by the growth of ice crystals, which means that extremely high porosity content can be achieved~\cite{Shao2016}. If this is not of particular interest for metals and ceramics--materials cannot be handled anymore when the porosity is too high--, this is extremely valuable for the development of aerogels, as discussed above. 
The second asset of ice-templating is that self-organization takes place locally, but everywhere in the bulk at the same time. Compared to most of the self-assembly processes, the average processing time for large samples (centimeters) is much lower. In addition, because the growing crystals segregate the particles locally, large defects are unlikely to develop. A direct consequence is the possibility to obtain a high reliability of the materials processed~\cite{Seuba2016b}. 
The combination of directionality and range of macropore size put ice-templated materials in a sweet spot with little competition. If many alternatives exist to make directional macropores, few of them present the same versatility and even fewer can hit a similar pore size. 
Another possibly unique asset is the availability of many processing parameters to adjust the architecture and microstructure of ice-templated materials and thus decouple, to some extent, structure and properties. However, because many processing parameters are interdependent, such adjustments require careful, exhaustive studies and a good understanding and control of the process~\cite{Miller2015,Seuba2016c}. 
Ice-templating is also mostly material-agnostic. The main requirement is a low solubility limit of the material to be templated in the crystal, to make sure that the growing crystals expel the material and that the particles are small enough to be repelled and concentrated. 
The entry barrier is low, little equipment is required. In the most extreme case of a solvent that sublimates at room temperature and atmospheric pressure--such as camphene-- and a material that does not need a high temperature treatment--a polymer--no equipment at all is required. It is possible to quickly master the basics of ice-templating and rapidly get started. 
Finally, we can combine ice-templating with other materials processing and shaping routes such as foaming~\cite{Lei2014}, tape-casting~\cite{Chen2014b}, spraying~\cite{Yu2013a}, extrusion~\cite{Moon2015}, co-extrusion~\cite{Moon2012}, or additive manufacturing~\cite{Adamkiewicz2015,Reed2016}. These combinations make it possible to achieve more complex, hierarchical architectures. 


\section{Roadmap} 
\label{sec:roadmap}

\begin{figure}[htbp]
\centering
\includegraphics[width=9cm]{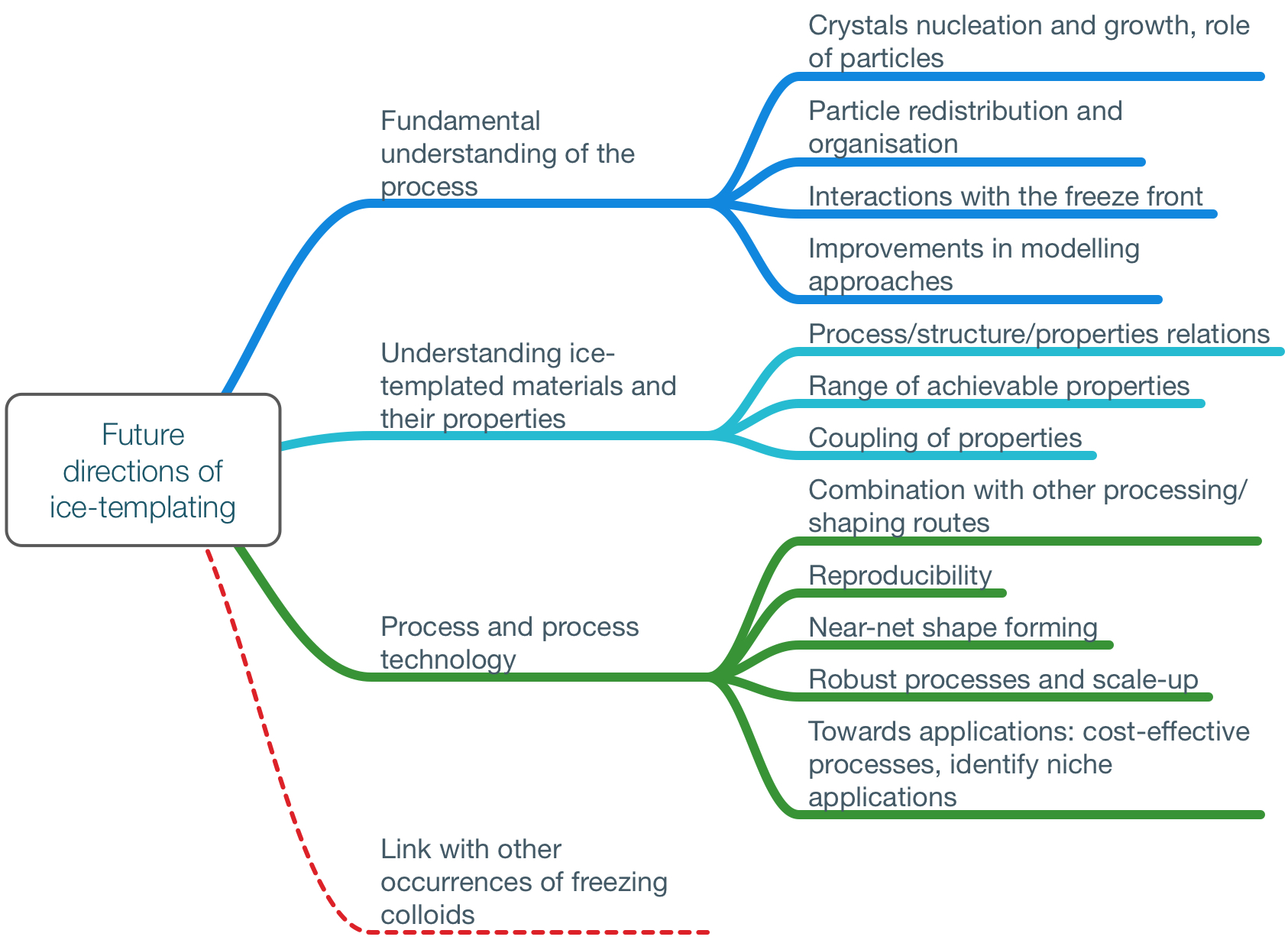}
\caption{Overview of the proposed roadmap for the future of ice-templating.}
\label{fig:roadmap}      
\end{figure}

Ice-templating has reached today the status of a well-established processing route. However, despite such a status and countless papers, only two commercial products can be found (a cellulose foam and an alumina foam, both for thermal insulation). I outline below three main domains that seems promising, either in terms of new science to unveil, or opportunities in terms of development and applications (Fig.~\ref{fig:roadmap}).

I want to emphasize first that progress in the domains discussed below will be facilitated by a multidisciplinary approach. The core of ice-templating is the interaction of objects with a freeze front, an ubiquitous phenomenon in nature and technology~\cite{Deville2017}. Linking ice-templating to these other occurrences will be a major facilitator both for the fundamental understanding and for the technological developments.

\subsection{Fundamental understanding of the process} 
\label{sub:fundamental_understanding_of_the_process}

Our understanding of the phenomena today is still only partial and very experimental~\cite{Miller2015}. Ice-templating is based on three core features: crystal growth, particle redistribution, and the interactions between the solidification fronts and objects. We now need to focus in particular on what occurs at the particle scale, and in particular the role of multiple particle interactions. Studies and analytical models developed so far focused almost exclusively on single-particle models~\cite{Rempel1999,Park2006}. We have now many evidences that we should question the validity of these models in concentrated systems. Understanding and predicting the behavior of ice crystals and of the particles in concentrated systems is essential to understand and control the crystals morphology, which eventually dictates the pores morphology and thus some of the properties. It is also important to understand the organization of particles between the crystals. The particle organization controls the homogeneity of the microstructure (will defects develop? can we sinter these materials properly?) but can also be used to develop particular textures~\cite{Bouville2014a} or architectures~\cite{Ferraro2016}. 
We badly need experimental observation of crystal growth and particle redistribution at particle scale. Although a lot of attention was dedicated to X-rays imaging~\cite{Delattre2014}, we are not yet able to use it to image such conditions (submicronic particle size, freeze front velocity in the 1--30~$\mu m/s$ range). The recent developments of confocal microscopy~\cite{Marcellini2016}, however, are an encouraging move in this direction. Optical microscopy, in particular in Hele-Shaw cells~\cite{You2016}, have also certainly an important role to play in a near future. Such experimental observations should help us for further developments of modelling approaches of freezing colloids. Phase field, in particular, has been developed for many years in materials science (essentially in metallurgy)~\cite{Aufgebauer2016}. A working phase field model of freezing suspensions would certainly be a major advance.


\subsection{Understanding ice-templated materials and their properties} 
\label{sub:understanding_ice_templated_materials_and_their_properties}

We are still far from a complete understanding of the process/structure relations. Most of the attention so far focused on the strength/porosity relations and even in this case, the studies are complex. One of the difficulty is that we need multiple descriptors of the structure to completely characterize the materials: we need to measure the pore size and morphology (which are anisotropic), but also the morphology of the walls comprising the pores (thickness, roughness, density, etc.). We should do this characterization in 3D as the structure of ice-templated materials may evolve along the freezing direction. We then need a set of well-defined architectures to assess the relative influence of each of these structural parameters. This is particularly tricky, as many of these parameters are interdependent. A good understanding of the process and its control--as well as a lot of patience--are required to perform such investigations. Knowing such relations, we should get a better idea of the assets and limitations of ice-templated materials. We can then compare them to alternatives processes and materials, a critical step before any industrial application can be developed. 

On a side note, the development of multifunctional ice-templated materials also raises questions regarding the coupling of properties in such materials and eventual size effects. Because we have many levers to control the size and characteristics of the porosity and architecture, ice-templating could also be an interesting playground to investigate such couplings.


\subsection{Process and process technology: towards commercial applications?} 
\label{sub:process_and_process_technology_towards_commercial_applications}

The third key of the future of ice-templating is the development of process technologies. Just two commercial ice-templated materials can be found today. If we want ice-templating to become a successful industrial process like tape-casting or extrusion, more technological work on the process will be required. Fortunately, we do not have to start from scratch. We can take inspiration and knowledge first from the other materials processing routes, but also from low temperature processes and technologies in other domains. Cost-effective techniques are already available at large scale in food industry~\cite{Goff2008} in particular and the process technology is already available to ensure a constant quality of the materials (ice cream) produced. There is thus no fundamental reason we could not ice-template in a cost-effective manner at large scale. 
The combination with other shaping and processing routes such as foaming or tape-casting not only provides additional levers to control the structure and properties of ice-templated materials, but may also facilitate the adoption of ice-templating. 
A key aspect not investigated so far is the reproducibility of the process (with the exception of~\cite{Naleway2016}). Ensuring predictable and reproducible structures and properties is essential for any industrial development. Recent approaches (such as the wedge technique~\cite{Pot2015}, which ensures a predictable control of the orientation of the ice-crystals perpendicular to the temperature gradient) should help, but more efforts are needed here.
One aspect was nearly not addressed so far: near-net shape forming. Almost all studies reported simple shapes (brick, cylinder). Such shapes could be suitable for a number of applications (such as catalysts supports or filtration), but for many applications, more complex shapes will be required. A few steps have been made in this direction, for example with the development of technologies to make ice-templated tubes with a radial morphology~\cite{Liu2013e,Seuba2016} for oxygen transport membranes applications. 
More work towards applications is thus needed. As with any new process, we need to focus on niche applications first, where the added value is clear and the cost less of an issue. A close collaboration with the industry will be essential for this transition.



\section{Conclusions} 
\label{sec:conclusions}

The segregation of particles by a freeze front is a ubiquitous phenomenon in nature and technology. Its use in ice-templating processes to obtain porous materials is one of the latest occurrences investigated and is thus relatively new. Despite being now a well-established topic, many opportunities await. We have now reached a stage where many proof of principle (in terms of materials, structure, properties) have been reported, but further work is required on the process and process technology developments. Despite being a simple process with a low entry barrier, ice-templating is an incredibly rich process. This is both an asset (as we have many levers to control the process, and thus the structure and properties of ice-templated materials) and an issue. Understanding the interdependent relations between all these parameters should still keep us busy for many years. A particular attention should be paid to the reproducibility and reliability of the associated processes and materials, which will determine if ice-templating may find practical applications or remain an academic topic of research, albeit a cool one.


\section*{Acknowledgements} 
\label{sec:acknowledgements}
 
The research leading to this paper has received funding from the European Research Council under the European Community's Seventh Framework Programme (FP7/2007-2013) Grant Agreement no. 278004, \textit{FreeCo}.


\section{References} 
\label{sec:references}

\bibliographystyle{Scripta-new-2.bst}
\bibliography{ice_templating_past_present_future}

\end{document}